\documentclass[sigconf,pbalance]{acmart}
\AtBeginDocument{%
  }

\copyrightyear{2026}
\acmYear{2026}
\setcopyright{cc}
\setcctype{by}
\acmConference[SIGIR-AP 2026]{Proceedings of the 2026 Annual International ACM SIGIR Conference on Research and Development in Information Retrieval in the Asia Pacific Region}{December 13--15, 2026}{Singapore, Singapore}
\acmBooktitle{Proceedings of the 2026 Annual International ACM SIGIR Conference on Research and Development in Information Retrieval in the Asia Pacific Region (SIGIR-AP 2026), December 13--15, 2026, Singapore, Singapore}
\acmDOI{10.1145/3848000.3850192}
\acmISBN{979-8-4007-3033-7/2026/12}

\usepackage{microtype}
\usepackage{url}
\usepackage{booktabs}
\usepackage{graphicx}
\usepackage{lineno}
\usepackage{enumitem}
\usepackage{listings} %
\usepackage{svg}
\usepackage{hyperref}

\definecolor{ darkblue}{rgb}{0, 0, 0.5}
\hypersetup{colorlinks=true, citecolor=darkblue, linkcolor=darkblue, urlcolor=darkblue}

\usepackage{multirow}
\usepackage{bigdelim}
\usepackage{todonotes}
\usepackage{longtable}
\usepackage{tabularray}
\usepackage{wrapfig}
\usepackage[most]{tcolorbox} 
\usepackage{url}
\usepackage{xspace}
\usepackage{svg}
\usepackage[absolute]{textpos} %

\usepackage{fdsymbol}   %

\usepackage[utf8]{inputenc} %
\usepackage[T1]{fontenc}    %
\usepackage{url}            %
\usepackage{booktabs}       %
\usepackage{nicefrac}       %
\usepackage{microtype}      %
\usepackage[table]{xcolor}         %
\definecolor{groupgray}{gray}{0.92}  
\usepackage[most]{tcolorbox}
\usepackage{csquotes}
\usepackage{wrapfig}
\usepackage{siunitx}
\usepackage{graphicx}
\usepackage{arydshln}
\usepackage{wrapfig}
\usepackage{enumitem}
\usepackage{soul} %

\usepackage{multirow}
\usepackage{xspace}
\usepackage{adjustbox}
\usepackage{pifont}
\usepackage{caption}
\usepackage{makecell}
\usepackage{subcaption}
\usepackage{bold-extra}
\usepackage{url}
\usepackage{float}
\usepackage{pgf-pie}
\usepackage{balance}
\makeatletter
\newcommand{\restorebodycolumns}{%
  \global\let\@outputdblcol\tevatron@normaloutputdblcol}
\makeatother
\usetikzlibrary{fit}

\usepackage[normalem]{ulem}

\newif\ifshowchanges
\showchangestrue

\usepackage{cleveref}
\crefname{figure}{Figure}{Figures}
\crefname{section}{Section}{Sections}
\crefname{equation}{Equation}{Equations}
\crefname{appendix}{Appendix}{Appendice}
\crefname{table}{Table}{Tables}

\definecolor{linkcolor}{RGB}{0, 0, 128}
\hypersetup{
     colorlinks   = true,
     citecolor    = linkcolor,
     linkcolor    = linkcolor,
     urlcolor     = linkcolor,
}
\usepackage{pifont}%
\newcommand{\cmark}{\ding{51}}%
\newcommand{\xmark}{\ding{55}}%
\usepackage{tikz}

\definecolor{codekw}{rgb}{0.13,0.13,0.55}
\definecolor{codestr}{rgb}{0.0,0.40,0.0}
\definecolor{codecmt}{rgb}{0.45,0.45,0.45}
\lstdefinestyle{tevpy}{
  language=Python,
  basicstyle=\ttfamily\scriptsize,
  keywordstyle=\color{codekw}\bfseries,
  stringstyle=\color{codestr},
  commentstyle=\color{codecmt},
  numbers=none,
  showstringspaces=false,
  breaklines=false,
  columns=fullflexible,
  keepspaces=true,
  frame=single,
  framerule=0.3pt,
  framesep=2pt,
  xleftmargin=2.3pt, xrightmargin=2.3pt,  
  rulecolor=\color{gray!50},
  aboveskip=0.5em, belowskip=0.5em,
}

\BeforeBeginEnvironment{lstlisting}{\par\nopagebreak\noindent\begin{minipage}{\linewidth}}
\AfterEndEnvironment{lstlisting}{\end{minipage}\par}

\usepackage{nicematrix}
\newcolumntype{L}[1]{>{\raggedright\let\newline\\\arraybackslash\hspace{0pt}}m{#1}}
\newcolumntype{C}[1]{>{\centering\let\newline\\\arraybackslash\hspace{0pt}}m{#1}}
\newcolumntype{R}[1]{>{\raggedleft\let\newline\\\arraybackslash\hspace{0pt}}m{#1}}
\newcolumntype{P}[1]{>{\centering\let\newline\\\arraybackslash\hspace{0pt}}m{#1}}
\newcommand{\sans}[1]{\textsf{#1}}

\newcommand{\huggingface}{\raisebox{-1.5pt}{\includegraphics[height=1.05em]{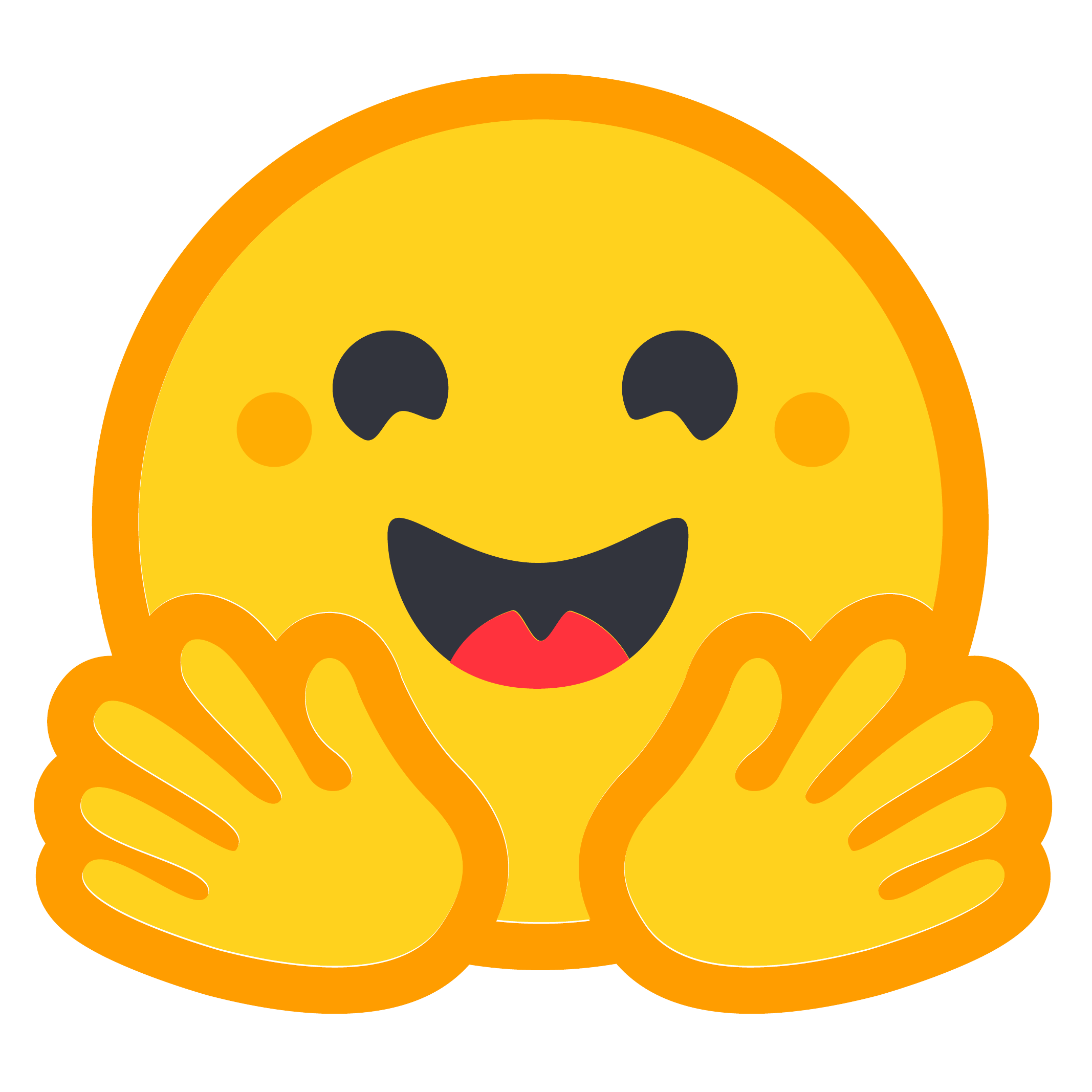}}\xspace}

\newcommand{\github}{\raisebox{-1.5pt}{\includegraphics[height=1.05em]{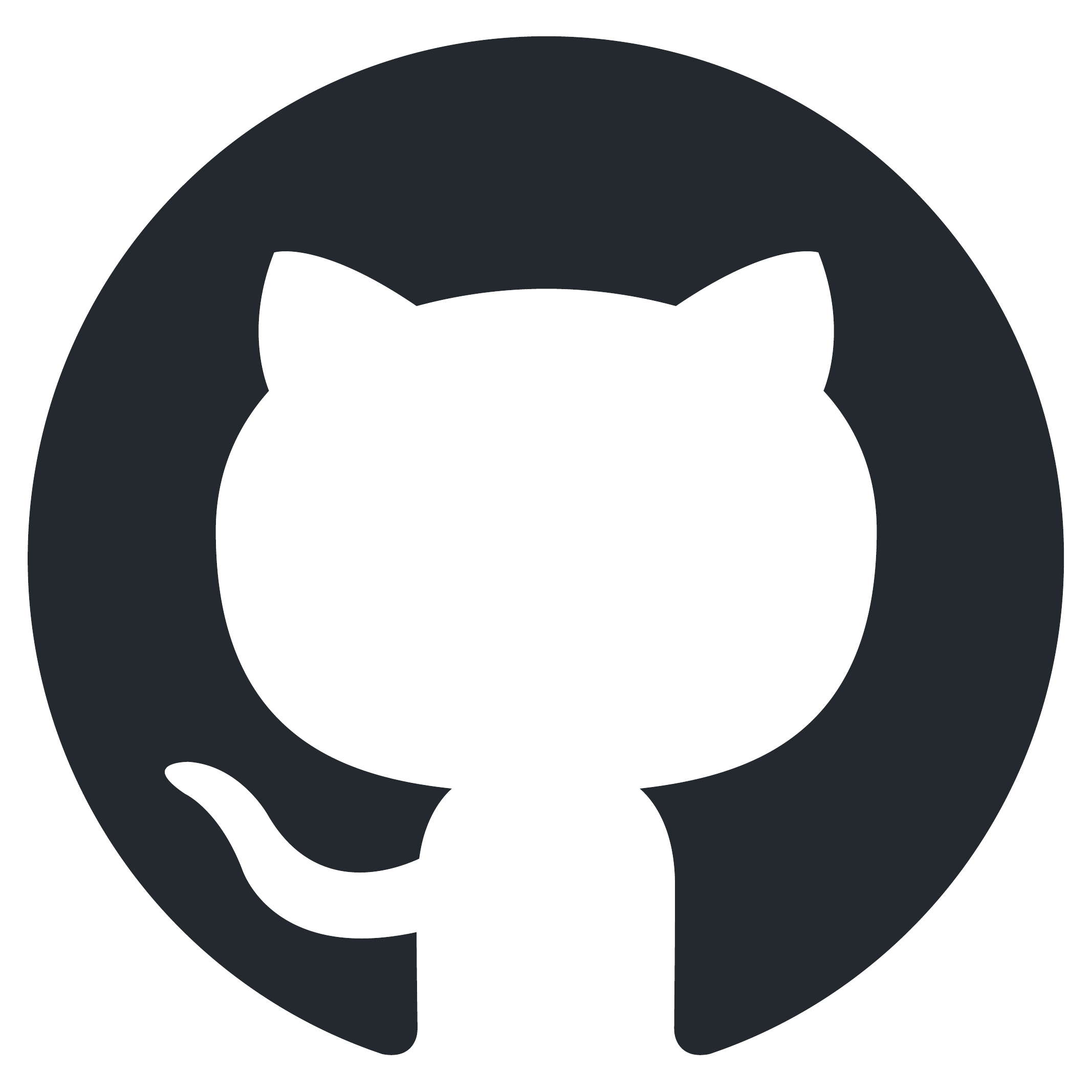}}\xspace}

\ccsdesc[500]{Information systems~Retrieval models and ranking}

\begin{document}

\title{Tevatron Meets Megatron: Expert-Parallel LLM Reranker Training on an Academic Budget}

\author{Zhichao Xu}
\email{zhichao.xu@utah.edu}
\affiliation{%
  \institution{University of Utah}
  \city{Salt Lake City}
  \state{UT}
  \country{USA}
}

\author{Xueguang Ma}
\email{x93ma@uwaterloo.ca}
\affiliation{%
  \institution{University of Waterloo}
  \city{Waterloo}
  \state{ON}
  \country{Canada}
}

\author{Shengyao Zhuang}
\email{s.zhuang@uq.edu.au}
\affiliation{%
  \institution{The University of Queensland}
  \city{Brisbane}
  \state{QLD}
  \country{Australia}
}

\author{Luyu Gao}
\email{luyug@cs.cmu.edu}
\affiliation{%
  \institution{Carnegie Mellon University}
  \city{Pittsburgh}
  \state{PA}
  \country{USA}
}

\author{Wenqian Ye}
\email{wenqian@virginia.edu}
\affiliation{%
  \institution{University of Virginia}
  \city{Charlottesville}
  \state{VA}
  \country{USA}
}

\author{Yu Wang}
\email{yuki.wang@utah.edu}
\affiliation{%
  \institution{University of Utah}
  \city{Salt Lake City}
  \state{UT}
  \country{USA}
}

\author{Jamie Callan}
\email{callan@cs.cmu.edu}
\affiliation{%
  \institution{Carnegie Mellon University}
  \city{Pittsburgh}
  \state{PA}
  \country{USA}
}

\author{Jimmy Lin}
\email{jimmylin@uwaterloo.ca}
\affiliation{%
  \institution{University of Waterloo}
  \city{Waterloo}
  \state{ON}
  \country{Canada}
}


\renewcommand{\shortauthors}{Zhichao Xu et al.}

\begin{abstract}
Training large dense and mixture-of-experts (MoE) rerankers is constrained by GPU memory and distributed-execution efficiency, which has kept academic reranker research largely on dense models. Tevatron trains rerankers through the Hugging Face Trainer with DeepSpeed and PyTorch FSDP1, a stack that stalls on three fronts: DeepSpeed ZeRO-2 has a known gradient-gathering failure on this path, FSDP1 full sharding is memory-safe but throughput-bound, and neither offers the expert parallelism (EP) that MoE training needs.
We present Tevatron~3.0, which integrates Megatron-Core as a reranker training backend while preserving Tevatron's data processing, evaluation pipeline, and Hugging Face checkpoint format.
For full-parameter dense 8B training, Megatron matches FSDP1 on BEIR-15 NDCG@10 with approximately 4\% lower epoch time and twice the peak GPU memory under matched data-parallel and recomputation settings, and approximately 22\% lower epoch time in its recommended configuration; it supports both LoRA and full-parameter fine-tuning.
Expert parallelism makes it possible to train a 30B-parameter MoE reranker (Qwen3-30B-A3B) on two nodes.
A controlled study of architecture, LoRA, and distillation on BEIR-15, with three first-stage retrievers and Hugging Face and vLLM serving, shows that the MoE reranker matches dense-8B quality at less than half the activated parameters and at higher inference throughput. We open-source the framework and trained checkpoints.\looseness=-1
\end{abstract}

\keywords{Tevatron, MoE Reranker, Megatron.}

\maketitle

\begin{center}
\small

\begin{minipage}{0.95\columnwidth}
\centering

\github~{\sans{Code:}}~
\href{https://github.com/texttron/tevatron}{\texttt{Tevatron v3}} \\[3pt]

\huggingface~{\sans{Models:}}~
\href{https://huggingface.co/collections/utahnlp/tevatron-v3}{Tevatron v3 Reranker Collection} \quad

\end{minipage}

\end{center}

\section{Introduction}
\label{sec:intro}

The Tevatron toolkit~\citep{gao2022tevatron} was first developed during the BERT~\citep{devlin-etal-2019-bert} era to give researchers
a flexible framework for training and evaluating neural retrievers and rerankers, and
Tevatron~2.0~\citep{ma2025tevatron20} extended it to billion-scale, multilingual, and multimodal dense
retrieval.
Reranking\,---\,re-scoring a first-stage candidate list with a more expensive
cross-encoder~\citep{xu2025surveymodelarchitecturesinformation}\,---\,has followed the same trajectory: state-of-the-art rerankers are
now billion-scale LLMs~\citep{zhang2025qwen3embedding}, increasingly trained with knowledge distillation against
strong teacher rerankers~\citep{hinton2015distillingknowledgeneuralnetwork}, and recent open-weight releases adopt
mixture-of-experts (MoE) backbones that activate only a fraction of their
parameters per token~\citep{nussbaum2025trainingsparsemixtureexperts,muennighoff2025generative}.
These recipes deliver strong quality, but reproducing them often requires specialized distributed-training engineering, which is unavailable to many academic research groups with limited engineering resources~\citep{shrestha2026mixtureofexpertslanguagemodelsstrong}.

Concretely, Tevatron's existing reranker training path\,---\,the Hugging Face Trainer,
whose memory-efficient training is provided by DeepSpeed~\citep{Rajbhandari2020deepspeed} and PyTorch FSDP1~\citep{zhao2023pytorchfsdpexperiencesscaling}\,---\,hits
three limitations on modern reranking workloads.
\begin{itemize}[leftmargin=*]
\item \textbf{DeepSpeed ZeRO-2} is less reliable on this path. A known
CPU-offloaded gradient accumulation bug makes ZeRO-2 sharding fail in practice,
forcing a fallback to PyTorch FSDP1.\footnote{Introduced in \url{https://github.com/deepspeedai/DeepSpeed/pull/6550} and not fixed until the recent \url{https://github.com/deepspeedai/DeepSpeed/pull/7967}.}
\item \textbf{PyTorch FSDP1 full-sharding} is memory-safe but
less efficient in terms of throughput. For an 8B reranker, plain data-parallel (DDP) runs out
of memory, while FSDP1 \texttt{full\_shard} (ZeRO-3) and \texttt{shard\_grad\_op} (ZeRO-2) both fit at near-identical wall-clock times\,---\,i.e., the heavier sharding stage
brings no speed improvement\,---\,and the distributed-optimizer backend delivers higher throughput than either FSDP1 configuration. 
\item \textbf{No expert parallelism.} 
PyTorch FSDP1 offers no expert parallelism,
so MoE rerankers (e.g., Qwen3-30B-A3B~\citep{yang2025qwen3technicalreport}, 128 experts) cannot be trained efficiently:
it can only replicate-then-shard the experts rather than route the active subset.
\end{itemize}

\noindent In this work we present Tevatron~3.0, which adds a Megatron training
backend~\citep{shoeybi2020megatronlmtrainingmultibillionparameter} to address these limitations while preserving the rest of the toolkit, together with a reworked evaluation interface.
Our contributions are:
\begin{itemize}[leftmargin=*]
\item \textbf{A Megatron reranker training backend} (C1) that is backward compatible
with Tevatron 2.0's data path, evaluation pipeline, and Hugging Face-loadable checkpoint
format, using bridge-based HF$\leftrightarrow$Megatron weight conversion. We
profile the existing FSDP settings on throughput and peak memory, then show that
the Megatron backend (ZeRO-1-style distributed optimizer with tensor parallelism)
matches FSDP-trained quality and\,---\,at matched data-parallel topology\,---\,matches its
efficiency, while its recommended configuration reduces dense-8B epoch time by $\sim$22\%,
supporting both LoRA~\citep{hu2021lora} and full-parameter fine-tuning as well as
listwise-KL distillation~\citep{bruch2019ananalysisofsoftmax,xu2025distillationversuscontrastivelearning} (a drop-in alternate trainer; \Cref{subsec:impl}). The toolkit also provides a unified evaluation
interface with two schemas\,---\,local scoring of existing ranklists, and
rerank-and-score against a remote HF/vLLM scoring-server pool over HTTP\,---\,behind
one CLI (detailed in~\Cref{subsec:eval-impl}).
\item \textbf{Expert-parallel MoE reranker training} (C2): Megatron's expert
parallelism trains a 30B-parameter MoE reranker on a single-job budget for which PyTorch FSDP1 offers no expert-parallel path, and we provide a named LoRA target-group registry so that low-rank adaptation can selectively target the appropriate expert layers in MoE models.
\item \textbf{A demonstration and released artifacts} (C3): to exercise the
system end-to-end, we evaluate combinations of model architecture (MoE 30B-A3B
vs.\ dense 8B), parameter efficiency (LoRA vs.\ full-parameter), and loss
formulation (listwise-KL distillation vs.\ contrastive), evaluated on BEIR-15~\citep{thakur2021beir}
with three first-stage retrievers, plus a serving-throughput matrix over
Hugging Face and vLLM backends~\citep{kwon2023efficientmemorymanagement}. The study is a capability demonstration rather
than the paper's central claim; we release all trained checkpoints as reusable
artifacts.
\end{itemize}

\noindent
The emphasis of this work is the system: a backend that makes these recipes
trainable and servable under an academic compute budget, with the data path,
evaluation, and checkpoint format unchanged.\footnote{Tevatron~3.0 also adds
other model classes to the toolkit\,---\,most notably decoder-LM learned sparse
retrieval (\textsc{LACONIC})~\cite{xu2026laconic,xu-etal-2025-csplade}, 
which turns a causal LM into a SPLADE-style first-stage retriever. 
We scope this paper to the reranker training backend and refer 
the reader to that work for the learned-sparse retriever.} 
We use a controlled setting to exercise the proposed system end-to-end\,---\,and, as one illustrative result, it shows \textit{the MoE reranker matching dense-8B quality at less than half the activated
parameters and higher inference throughput}, a favorable operating point that the
framework now makes straightforward to reproduce.

\section{Tevatron 3.0 System Design}
\label{sec:backend}

Instead of an overhaul, Tevatron~3.0 is a set of \emph{additive} features on top of Tevatron 2.0. 
The guiding principle is \textit{backward compatibility}: every new capability
below\,---\,the Megatron training backend~\citep{shoeybi2020megatronlmtrainingmultibillionparameter}, distillation, and the dual-schema
evaluation interface\,---\,plugs into the existing framework's data path, collator,
and checkpoint conventions, reusing them unchanged. A practitioner can easily launch new experiments by changing a launch script or adding a flag, instead of migrating their data or
re-learning the toolkit.

We first describe
the design principle~(\S\ref{subsec:design-principle}), the implementation of the core training backend~(\S\ref{subsec:impl}), then the evaluation layer~(\S\ref{subsec:eval-impl}), the extension
points exposed to toolkit users~(\S\ref{subsec:extensibility}), and finally the characterization of backends~(\S\ref{subsec:profiling}).

\subsection{Design Principle}
\label{subsec:design-principle}

The most significant addition of Tevatron 3.0 is a Megatron-Core training backend alongside the
existing Hugging Face Trainer path, exposed through the same command-line interface
and consuming the same training data. The design goal is \emph{substitutability}:
a user switches backends by changing a launch script, instead of their data, their
evaluation, or the delivery format of the checkpoint. Three properties make this
possible.

First, the backend reuses the toolkit's existing data path and evaluation
pipeline without modification. The dataset, collator, and BEIR~\citep{thakur2021beir} evaluation modules are shared with
the Hugging Face path; only the trainer object differs. A reranker is scored as a
standard causal language model\,---\,the log-odds of the ``yes''/``no'' answer tokens
at the final prompt position\,---\,so no architecture-specific value head is
introduced and the same scoring code runs under any parallelism configuration
(\Cref{subsec:impl}).

\begin{table}[t!]
\centering
\caption{Functionality comparison between the existing PyTorch FSDP1 backend
(via the Hugging Face Trainer) and the proposed Megatron backend. 
\textsuperscript{$\dagger$}Parameter/gradient sharding is currently achieved
through tensor parallelism; direct FSDP-style ZeRO-2/3 data-parallel sharding is
left to future work. The PyTorch FSDP1 column describes the reranker training backend used in Tevatron~2.0.}
\footnotesize
\setlength{\tabcolsep}{4pt}
\begin{tabular}{lcc}
\toprule
Capability & PyTorch FSDP1 & Megatron \\
\midrule
Data parallelism (DP)              & \cmark & \cmark \\
Optimizer-state sharding (ZeRO-1)     & \cmark & \cmark \\
Grad/param sharding (ZeRO-2/3) & \cmark & via TP\textsuperscript{$\dagger$} \\
Tensor parallelism (TP)            & \xmark & \cmark \\
Pipeline parallelism (PP)          & \xmark & \cmark \\
Expert parallelism (EP, MoE) & \xmark & \cmark \\
Activation recompute               & \cmark & \cmark \\
Full-parameter fine-tuning         & \cmark & \cmark \\
LoRA fine-tuning                   & \cmark & \cmark \\
MoE-aware LoRA target groups       & \xmark & \cmark \\
Reuses Tevatron data + eval        & \cmark & \cmark \\
HF-format checkpoints (vLLM-ready) & \cmark & \cmark \\
\bottomrule
\end{tabular}

\label{tab:feature-matrix}
\end{table}

%
%
%
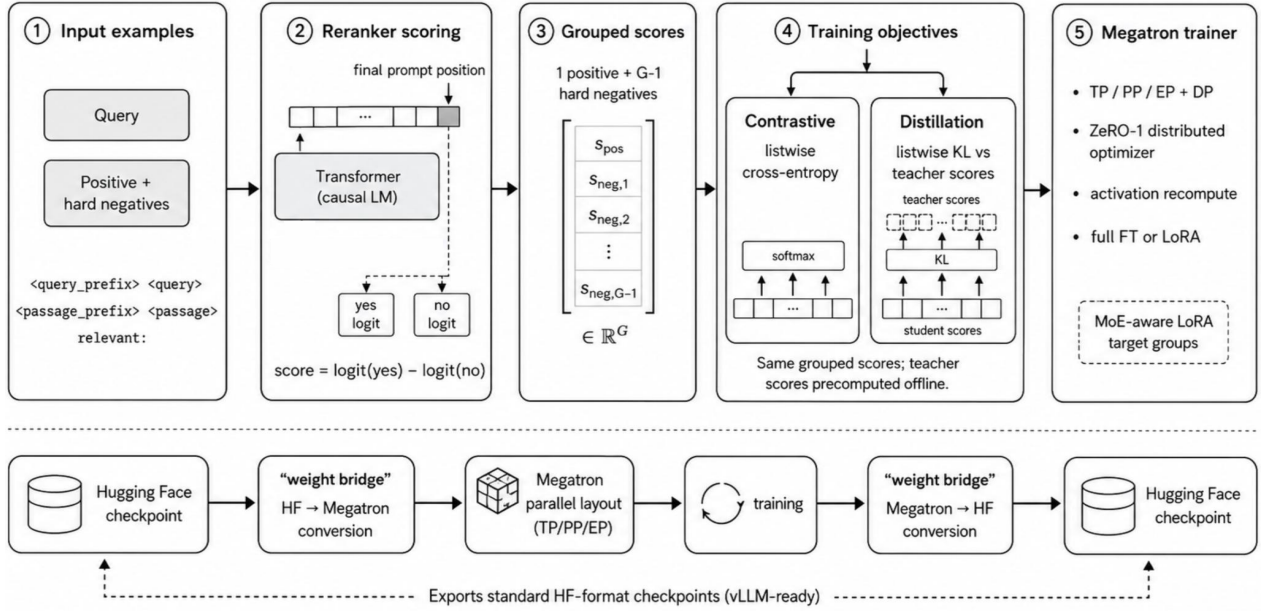
\begin{figure*}[t]
\centering
\definecolor{inFill}{HTML}{DCE9F5}\definecolor{inLine}{HTML}{3B6EA5}\definecolor{inDark}{HTML}{9DBFE0}
\definecolor{mdFill}{HTML}{FBE3CC}\definecolor{mdLine}{HTML}{C2691F}
\definecolor{scFill}{HTML}{DDEFD9}\definecolor{scLine}{HTML}{4C8C3F}\definecolor{scDark}{HTML}{A9D3A0}
\definecolor{tcFill}{HTML}{E8E0F0}\definecolor{tcLine}{HTML}{7A5C9E}
\begin{tikzpicture}[
  font=\footnotesize,
  line width=0.5pt,
  >=stealth,
  panel/.style={draw, rounded corners=3pt},
  box/.style={draw, rounded corners=1.5pt, align=center, inner sep=2pt},
  inp/.style={fill=inFill, draw=inLine},
  model/.style={fill=mdFill, draw=mdLine},
  score/.style={fill=scFill, draw=scLine},
  num/.style={draw, circle, inner sep=0pt, minimum size=0.32cm, font=\scriptsize\bfseries},
  title/.style={anchor=west, font=\footnotesize\bfseries},
  note/.style={font=\scriptsize, align=center},
]

\draw[panel] (0,0)     rectangle (3.6,4.8);
\draw[panel] (4.0,0)   rectangle (7.2,4.8);
\draw[panel] (7.6,0)   rectangle (10.1,4.8);
\draw[panel] (10.5,0)  rectangle (14.3,4.8);
\draw[panel] (14.7,0)  rectangle (17.6,4.8);

\foreach \a/\b in {3.6/4.0, 7.2/7.6, 10.1/10.5, 14.3/14.7}
  \draw[->, line width=0.7pt] (\a,2.4) -- (\b,2.4);

\node[num] at (0.30,4.5) {1};
\node[title] at (0.50,4.5) {Input examples};
\node[box, inp, minimum width=2.4cm, minimum height=0.55cm] at (1.8,3.65) {Query};
\node[box, inp, minimum width=2.4cm, minimum height=0.8cm]  at (1.8,2.65) {Positive +\\hard negatives};
\node[font=\scriptsize\ttfamily, align=center] at (1.8,1.1)
  {<query\_prefix> <query>\\[2pt]<passage\_prefix> <passage>\\[2pt]relevant:};

\node[num] at (4.30,4.5) {2};
\node[title] at (4.50,4.5) {Reranker scoring};
\foreach \i in {0,...,6}
  \draw[fill=inFill] (4.3+0.35*\i,3.275) rectangle ++(0.35,0.35);
\draw[fill=inDark] (6.40,3.275) rectangle (6.75,3.625);
\node at (5.525,3.45) {$\cdots$};
\node[note, anchor=south east] at (6.85,3.93) {final prompt position};
\draw[->] (6.575,3.93) -- (6.575,3.65);
\node[box, model, minimum width=1.9cm, minimum height=0.75cm] at (5.1,2.45) {Transformer\\(causal LM)};
\draw[->] (4.55,2.825) -- (4.55,3.275);
\node[box, score, minimum width=0.95cm, minimum height=0.6cm, font=\scriptsize] (yes) at (5.2,1.25)   {yes\\logit};
\node[box, score, minimum width=0.95cm, minimum height=0.6cm, font=\scriptsize] (no)  at (6.575,1.25) {no\\logit};
\draw[->, dashed] (6.575,3.275) -- (no.north);
\draw[->, dashed] (6.575,1.95) -| (yes.north);
\node[note] at (5.6,0.42) {$\mathrm{score} = \mathrm{logit}(\texttt{yes}) - \mathrm{logit}(\texttt{no})$};

\node[num] at (7.90,4.5) {3};
\node[title] at (8.10,4.5) {Grouped scores};
\node[note] at (8.85,3.85) {1 positive +\\$G{-}1$ hard negatives};
\foreach \lab/\f [count=\i from 0] in
    {$s_{\mathrm{pos}}$/scDark, $s_{\mathrm{neg},1}$/scFill, $s_{\mathrm{neg},2}$/scFill, $\vdots$/scFill, $s_{\mathrm{neg},G-1}$/scFill} {
  \draw[fill=\f] (8.35,3.35-0.38*\i) rectangle (9.35,2.97-0.38*\i);
  \node at (8.85,3.16-0.38*\i) {\lab};
}
\draw (8.30,3.40) -- (8.22,3.40) -- (8.22,1.40) -- (8.30,1.40);
\draw (9.40,3.40) -- (9.48,3.40) -- (9.48,1.40) -- (9.40,1.40);
\node at (8.85,0.95) {$\in \mathbb{R}^{G}$};

\node[num] at (10.80,4.5) {4};
\node[title] at (11.00,4.5) {Training objectives};
\draw[rounded corners=2pt] (10.65,0.85) rectangle (12.30,4.05);
\draw[rounded corners=2pt] (12.50,0.85) rectangle (14.15,4.05);
\node[font=\footnotesize\bfseries] at (11.475,3.80) {Contrastive};
\node[note] at (11.475,3.30) {listwise\\cross-entropy};
\node[box, minimum width=1.2cm, minimum height=0.28cm, inner sep=1pt, font=\scriptsize] at (11.475,1.95) {softmax};
\foreach \i in {0,...,4}
  \draw[fill=scFill] (10.825+0.26*\i,1.25) rectangle ++(0.26,0.22);
\node[font=\tiny] at (11.475,1.36) {$\cdots$};
\foreach \x in {10.955,11.475,11.995}
  \draw[->] (\x,1.47) -- (\x,1.81);
\node[font=\tiny] at (11.475,1.08) {scores};
\node[font=\footnotesize\bfseries] at (13.325,3.80) {Distillation};
\node[note] at (13.325,3.30) {listwise KL vs\\teacher scores};
\node[font=\tiny] at (13.325,2.72) {teacher scores};
\foreach \i in {0,...,4}
  \draw[fill=tcFill] (12.675+0.26*\i,2.35) rectangle ++(0.26,0.22);
\node[font=\tiny] at (13.325,2.46) {$\cdots$};
\node[box, minimum width=1.2cm, minimum height=0.28cm, inner sep=1pt, font=\scriptsize] at (13.325,1.95) {KL};
\foreach \x in {12.805,13.325,13.845} {
  \draw[->] (\x,2.35) -- (\x,2.09);
  \draw[->] (\x,1.47) -- (\x,1.81);
}
\foreach \i in {0,...,4}
  \draw[fill=scFill] (12.675+0.26*\i,1.25) rectangle ++(0.26,0.22);
\node[font=\tiny] at (13.325,1.36) {$\cdots$};
\node[font=\tiny] at (13.325,1.08) {student scores};
\node[note, text width=3.6cm] at (12.4,0.42) {Same grouped scores; teacher scores precomputed offline.};

\node[num] at (14.95,4.5) {5};
\node[title] at (15.13,4.5) {Megatron trainer};
\node[anchor=north west, inner sep=0pt] at (14.85,4.0) {%
  \begin{tabular}{@{}l@{\ }p{2.3cm}@{}}
    \textbullet & TP / PP / EP + DP\\[4pt]
    \textbullet & \raggedright ZeRO-1 distributed optimizer\tabularnewline[4pt]
    \textbullet & activation recompute\\[4pt]
    \textbullet & full FT or LoRA
  \end{tabular}};
\node[box, dashed, minimum width=2.2cm, minimum height=0.75cm] at (16.15,0.8) {MoE-aware LoRA\\target groups};

\draw[dashed, gray] (0,-0.4) -- (17.6,-0.4);

\begin{scope}[every node/.style={box, text width=2.42cm, minimum height=1.05cm}]
  \node[inp] (b1) at (1.28,-1.35) {Hugging Face\\checkpoint};
  \node (b2) at (4.29,-1.35) {\textbf{Weight bridge}\\HF $\to$ Megatron\\conversion};
  \node[model] (b3) at (7.30,-1.35) {Megatron parallel\\layout (TP/PP/EP)};
  \node[model] (b4) at (10.31,-1.35) {Training};
  \node (b5) at (13.32,-1.35) {\textbf{Weight bridge}\\Megatron $\to$ HF\\conversion};
  \node[inp] (b6) at (16.33,-1.35) {Hugging Face\\checkpoint};
\end{scope}
\foreach \u/\v in {b1/b2, b2/b3, b3/b4, b4/b5, b5/b6}
  \draw[->, line width=0.7pt] (\u) -- (\v);

\node[note] (exp) at (8.8,-2.45) {Exports standard HF-format checkpoints (vLLM-ready)};
\draw[->, dashed] (exp.west) -| (b1.south);
\draw[->, dashed] (exp.east) -| (b6.south);

\end{tikzpicture}
\caption{Overview of the Megatron training backend. Query--passage
    examples are scored by a standard causal language model and organized
    into listwise groups for contrastive or distillation training. The
    Megatron backend provides tensor, pipeline, and expert parallelism while
    preserving Hugging Face checkpoint compatibility through a bidirectional
    weight bridge.}
\Description{Query--passage examples are scored by a Hugging Face causal language model, organized into listwise groups, and processed by the Megatron trainer. The trainer supports tensor, pipeline, and expert parallelism, a distributed optimizer, activation recomputation, full fine-tuning, and LoRA. A bidirectional weight bridge converts between Hugging Face checkpoints and the Megatron parallel layout.}
\label{fig:backend_overview}
\end{figure*}

Second, weights cross the Hugging Face~$\leftrightarrow$~Megatron boundary in both directions through a
weight bridge:\footnote{\url{https://docs.nvidia.com/nemo/megatron-bridge/latest/}} a pretrained Hugging Face checkpoint is
sharded into the Megatron parallel layout at load time, and the trained model is
exported back to Hugging Face format at save time. Full-parameter fine-tuning and LoRA use
the corresponding bridge save paths, while the latter writes pre-merged weights. The
result is that every Tevatron~3.0 checkpoint loads through the ordinary
Hugging Face and vLLM paths with no conversion step, preserving the rest of the
ecosystem.

Third, the backend exposes the parallelism dimensions that
PyTorch FSDP1 lacks: tensor (TP), pipeline (PP), and expert (EP) model parallelism. Data parallelism (DP) spans the remaining ranks after TP and PP are configured, while EP partitions the experts of each MoE layer across the ranks within a data-parallel group; in the configurations we use, EP divides DP rather than adding a further multiplicative dimension (\cref{sec:moe}). Megatron’s distributed optimizer shards optimizer states across DP ranks while replicating parameters and gradients, corresponding to a ZeRO-1-style scheme. This design minimizes communication overhead and, as shown in \cref{subsec:profiling}, is faster than the FSDP1 configurations used by the Hugging Face Trainer in its recommended topology and within a few percent at matched topology, in exchange for higher per-rank memory at the same data-parallel degree, because ZeRO-1 replicates parameters and gradients where FSDP1 shards them. The backend's memory levers beyond ZeRO-1 optimizer-state sharding are therefore tensor parallelism for dense layers and expert parallelism for MoE experts, not FSDP-style parameter and gradient sharding. Critically, EP unblocks MoE reranker training. We treat the precise TP/PP/EP configuration as a hardware-dependent deployment choice; the framework documentation provides recommended configurations for different model scales.

\Cref{tab:feature-matrix} summarizes the functionality gap between the two
backends. 

The Megatron backend adds tensor, pipeline, and expert parallelism alongside support for activation recomputation. It retains backward compatibility with Tevatron’s existing data formats, evaluation pipeline, and Hugging Face checkpoints.

\subsection{Implementation}
\label{subsec:impl}
\Cref{fig:backend_overview} summarizes the training path. The Megatron
backend reuses Tevatron's existing dataset, collator, and evaluation
components; the trainer and distributed model execution are the only
backend-specific components. The implementation follows the same sequence
under both contrastive and distillation training: query--passage examples are
converted into generative relevance scores, those scores are arranged into
listwise groups, an objective is applied to each group, and gradients are
updated under the configured parallel topology. The resulting checkpoint is
then exported back to Hugging Face format through the weight bridge.

\paragraph{Reranker scoring.}
We format each query--passage pair as
\begin{center}\footnotesize\texttt{<query\_prefix><query> <passage\_prefix><passage> relevant:}\end{center}
and score it with a generative reranker as the difference between the
``yes'' and ``no'' answer-token logits at the final non-padding prompt position. We define a scalar relevance score:
\[
s_\theta(q,p)
=
z_\theta(q,p)_{\mathrm{yes}}
-
z_\theta(q,p)_{\mathrm{no}}.
\]
For each query, one positive passage and $G{-}1$ hard negatives are scored
independently and then reshaped into a grouped score vector
$\mathbf{s}_i\in\mathbb{R}^{G}$. Because scoring uses only the ordinary causal
LM output tensor, the same formulation applies without modification under
data, tensor, pipeline, and expert parallelism.

\begin{lstlisting}[style=tevpy]
# output_tensor: (B, S, V) LM logits; attention_mask: (B, S)
seq_lengths = attention_mask.sum(dim=-1) - 1  # last real token
last_logits = output_tensor[torch.arange(B), seq_lengths]   # (B, V)
scores = last_logits[:, yes_token_id] - last_logits[:, no_token_id]
grouped_scores = scores.view(-1, G)  # (n_groups, G); positive: 0
\end{lstlisting}

\paragraph{Contrastive and distillation losses.}

Both training objectives operate on the same grouped scores $\mathbf{s}$. For
contrastive training, the positive passage is placed at a fixed position in
each group and optimized with listwise cross-entropy. For distillation, the
student distribution over the same $G$ passages is matched to a softened
teacher distribution using listwise KL divergence. Teacher scores are computed
offline and stored with the training examples, so distillation introduces no
teacher model or additional forward pass during training. It therefore changes
only the loss applied to the grouped scores~$\mathbf{s}$; the data path, scoring logic,
parallel configuration, and checkpoint procedure remain unchanged.
\begin{lstlisting}[style=tevpy]
if loss_kind == "contrastive":
    labels = torch.zeros(grouped_scores.size(0), dtype=torch.long)
    loss = F.cross_entropy(grouped_scores, labels)  # positive: 0
elif loss_kind == "distill":
    student = F.log_softmax(grouped_scores / s_temp, dim=-1)
    teacher = F.softmax(teacher_scores.view(-1, G) / t_temp, dim=-1)
    loss = F.kl_div(
        student, teacher, reduction="batchmean") * (s_temp ** 2)
\end{lstlisting}

\paragraph{Parallelism and LoRA}
Parallelism is critical in scaling up the model size in a multi-GPU, multi-node distributed training setup. In Tevatron~3.0, parallelism is configured by three flags: tensor (TP), pipeline
(PP), and expert (EP) model-parallel sizes, with data
parallelism derived as $\mathrm{DP} = \text{world\_size} / (\mathrm{TP}\cdot\mathrm{PP})$.
Expert parallelism is carved out of the data-parallel group, with $\mathrm{EP}$ dividing $\mathrm{DP}$ in the configurations we use (\cref{sec:moe}). The data-parallel scheme uses Megatron's
distributed optimizer (ZeRO-1-style: optimizer state sharded across DP ranks,
parameters and gradients replicated). LoRA is injected through a pre-wrap hook so
adapter parameters land in the DDP gradient buffers and the distributed optimizer
sees only the trainable adapters. For MoE models, a named target-group registry
maps roles to module patterns, preventing naive dense patterns from inadvertently matching all the expert layers.
\begin{lstlisting}[style=tevpy]
LORA_TARGET_GROUPS = {
    "attn":        ("linear_qkv", "linear_proj"),
    "mlp":         ("linear_fc1", "linear_fc2"),  # dense MLP only
    "moe_experts": (
        "*.experts.*.linear_fc1", "*.experts.*.linear_fc2"),
    "moe_shared":  ("*.shared_experts.linear_fc1",
                    "*.shared_experts.linear_fc2"),
    "moe_router":  ("*.router.weight",),  # usually frozen
}
\end{lstlisting}

\paragraph{Weight bridge.} A pretrained Hugging Face model is sharded into the Megatron parallel layout at the start of the distributed training. When checkpointing during or after the training run, the model in Megatron parallel layout is written to disk in Hugging Face format via the weight bridge (full parameters for full fine-tuning; pre-merged weights for LoRA), so they load
through the standard Hugging Face and vLLM paths with no conversion step.

\paragraph{Distillation training.}
Distillation uses a sibling training driver with the same parallelism flags.
Teacher scores are annotated \emph{offline} by
\texttt{tevatron.utils.annotate\_with\_teacher}, then loaded by the trainer,
which exposes the teacher and student temperatures. The scoring and checkpoint
paths are shared with contrastive training.
\begin{lstlisting}[style=tevpy]
# Contrastive (listwise CE):
torchrun ... -m tevatron.megatron.driver.train \
    --model_name_or_path Qwen3-8B-Base --dataset_name rlhn/rlhn-680K \
    --train_group_size 8 ...

# Distillation (listwise KL; offline teacher scores).
# Same parallelism flags; add dataset and temperatures.
torchrun ... -m tevatron.megatron.driver.distill_train \
    --model_name_or_path Qwen3-8B-Base \
    --distill_dataset_path rlhn-680K-qwen3-reranker-8b-top200 \
    --train_group_size 8 --teacher_temp 2.0 --student_temp 1.0 ...
\end{lstlisting}

\subsection{A Unified Evaluation and Serving Pipeline}
\label{subsec:eval-impl}

%
%
%
\begin{figure}[t]
\centering
\definecolor{inFill}{HTML}{DCE9F5}\definecolor{inLine}{HTML}{3B6EA5}\definecolor{inDark}{HTML}{9DBFE0}
\definecolor{mdFill}{HTML}{FBE3CC}\definecolor{mdLine}{HTML}{C2691F}
\definecolor{scFill}{HTML}{DDEFD9}\definecolor{scLine}{HTML}{4C8C3F}\definecolor{scDark}{HTML}{A9D3A0}
\definecolor{tcFill}{HTML}{E8E0F0}\definecolor{tcLine}{HTML}{7A5C9E}
\begin{tikzpicture}[
  yscale=1.16,
  font=\scriptsize,
  line width=0.5pt,
  >=stealth,
  box/.style={draw, rounded corners=1.5pt, align=center, inner sep=2.5pt},
  group/.style={draw, rounded corners=3pt, inner sep=4pt},
  head/.style={align=center, inner sep=1pt},
  inp/.style={fill=inFill, draw=inLine},
  model/.style={fill=mdFill, draw=mdLine},
  score/.style={fill=scFill, draw=scLine},
]

\begin{scope}[every node/.append style={box, text width=2.2cm, minimum height=1.05cm}]
  \node (reg) at (1.328,0) {\textbf{Benchmark registry}\\BEIR-15, MTEB,\\BRIGHT, private set};
  \node (met) at (4.104,0) {\textbf{Metric core}\\NDCG@$k$, Recall@$k$,\\MAP, MRR};
  \node (tab) at (6.880,0) {\textbf{Table renderer}\\aggregate +\\summary tables};
\end{scope}
\node[head, anchor=south] (blt) at (4.104,0.56) {\footnotesize\textbf{Benchmark layer}};
\node[group, fit=(reg)(met)(tab)(blt)] (bl) {};
\draw[->] (reg) -- (met);
\draw[->] (met) -- (tab);

\node[box, model, text width=5.6cm, minimum height=1.35cm] (ev) at (4.104,-1.8) {%
  {\footnotesize\textbf{Unified evaluator}}\\[1pt]
  \texttt{tevatron.eval.run}\\[2pt]
  \begin{tabular}{@{}l@{\hspace{1em}}l@{}}
    \textbullet~benchmark evaluation  & \textbullet~ranklist aggregation\\
    \textbullet~dataset orchestration & \textbullet~shared CLI
  \end{tabular}};
\draw[->] (bl.south) -- (ev.north);

\begin{scope}[yshift=-0.45cm]
\node[head, anchor=north] (s1t) at (1.253,-2.95) {\footnotesize\textbf{Schema 1}\\score-only (local)};
\begin{scope}[every node/.append style={box, text width=2.05cm, minimum height=0.6cm}]
  \node[inp] (rl)  at (1.253,-3.90) {Existing\\ranklists};
  \node (sa)  at (1.253,-5.60) {score + aggregate\\\textit{no model load}};
  \node[inp] (rt1) at (1.253,-7.30) {results table};
\end{scope}
\draw[->] (rl) -- (sa);
\draw[->] (sa) -- (rt1);
\node[group, fit=(s1t)(rl)(sa)(rt1)] (s1) {};

\node[head, anchor=north] (s2t) at (5.515,-2.95) {\footnotesize\textbf{Schema 2}\\rerank-and-score (remote)};
\begin{scope}[every node/.append style={box, text width=2.0cm, minimum height=0.6cm}]
  \node[inp] (cb)   at (4.05,-3.90) {Candidate batches};
  \node (cl)   at (4.05,-4.75) {HTTP \texttt{/score}\\client};
  \node (pool) at (4.05,-5.60) {backend pool};
  \node[inp] (rr)   at (4.05,-6.45) {reranked\\ranklists};
  \node[inp] (rt2)  at (4.05,-7.30) {results table};
  \node[score] (hf)   at (6.88,-5.18) {\textbf{HF server}\\single-GPU replicas};
  \node[score] (vl)   at (6.88,-6.02) {\textbf{vLLM server}\\tensor-parallel};
\end{scope}
\node[head] (con) at (6.88,-6.60) {identical \texttt{/score} contract};
\node[draw, dashed, rounded corners=2pt, inner sep=3pt, fit=(hf)(vl)(con)] {};
\draw[->] (cb) -- (cl);
\draw[->] (cl) -- (pool);
\draw[->] (pool) -- (rr);
\draw[->] (rr) -- (rt2);
\draw[dashed] (pool.east) -- ++(0.25,0) coordinate (j);
\draw[->, dashed] (j) |- (hf.west);
\draw[->, dashed] (j) |- (vl.west);
\node[group, fit=(s2t)(cb)(rt2)(hf)(vl)(con)] (s2) {};

\end{scope}

\draw (ev.south) -- ++(0,-0.45) coordinate (k);
\draw[->] (k) -| (s1.north);
\draw[->] (k) -| (s2.north);

\end{tikzpicture}
\caption{Overview of the unified evaluation and serving pipeline. The same
    evaluator supports local scoring of existing ranklists and remote
    rerank-and-score evaluation through interchangeable Hugging Face and vLLM
    serving.}
\Description{A benchmark layer provides datasets, metrics, and table rendering to a unified evaluator. The evaluator either scores existing ranklists locally or sends candidate batches through an HTTP scoring client to Hugging Face single-GPU replicas or tensor-parallel vLLM servers, then aggregates the reranked results.}
\label{fig:evaluation}
\end{figure}
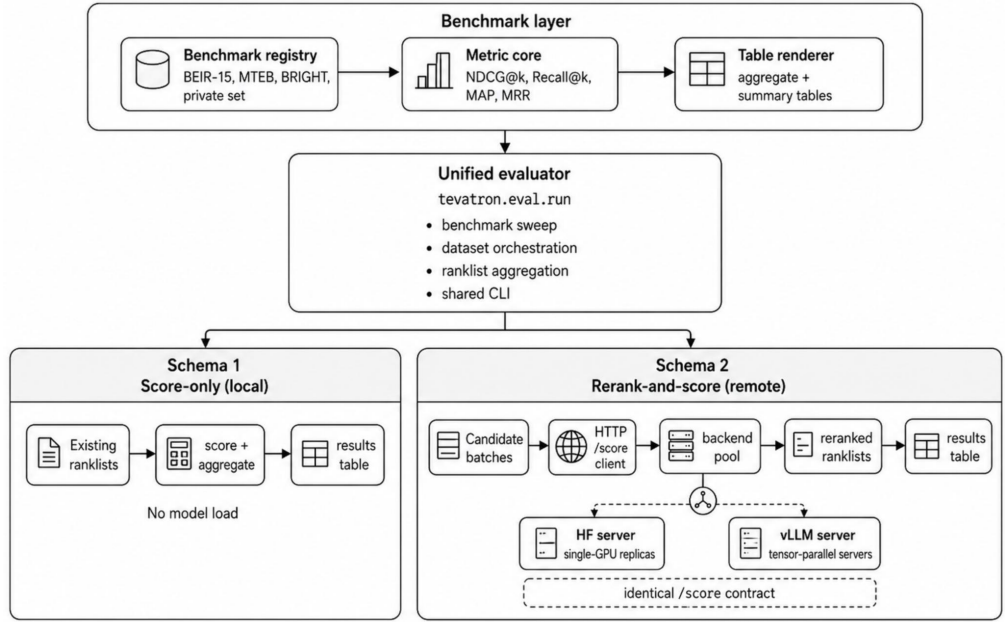

Evaluation is likewise an additive layer over the existing reranking-eval code
rather than a replacement. It separates three concerns\,---\,an invariant metric
core (\texttt{score(qrels, results, k)} $\rightarrow$ NDCG@$k$ / Recall@$k$ /
MAP / MRR), a benchmark registry (which datasets, where qrels come from, how
they aggregate), and a generic table renderer\,---\,so a benchmark beyond BEIR-15 is
a registry entry rather than a code change (\Cref{subsec:extensibility}). As shown in~\autoref{fig:evaluation}, a unified
entry point (\texttt{tevatron.eval.run}) runs benchmark evaluation\,---\,aggregating
per-dataset scores into the BEIR-15 mean\,---\,and supports two usage schemas behind
the same CLI:

\begin{enumerate}[leftmargin=1.4em,itemsep=0.5\baselineskip]
\item \textbf{Score-only (local).} Given existing ranklists (first-stage runs or
already-reranked outputs), it scores and aggregates directly without loading a model:
\begin{lstlisting}[style=tevpy]
python -m tevatron.eval.run \
    --ranklist_pattern /.../{dataset}/rank.text \
    --results_dir results/e5-base --name e5-base-v2
\end{lstlisting}

\item \textbf{Rerank-and-score via a remote HTTP pool.}\newline The reranker
is hosted as a persistent scoring server\,---\,\texttt{vllm} or \texttt{hf}
backend\,---\,behind a small HTTP API (\texttt{POST /score}); the evaluator becomes a
thin client that dispatches candidate batches to one or more backend URLs and
reassembles the ranklists. The model is loaded once per pool and reused throughout
the evaluation, and the pool can run on a remote (multi-GPU) node while the
evaluation is driven from elsewhere:
\begin{lstlisting}[style=tevpy]
# Host on a remote GPU node; load the model once.
python -m tevatron.eval.serve.server --backend vllm \
    --model <ckpt> --tensor_parallel_size 2 --port 8100

# Client (anywhere): rerank+score against one or more backend URLs.
python -m tevatron.eval.run \
    --backends http://node:8100 http://node:8101 ... \
    --rerank_input_pattern /.../{dataset}/rerank.jsonl \
    --results_dir results/<ckpt>
\end{lstlisting}
\end{enumerate}

\noindent Both backends expose the identical \texttt{/score} contract, so the same client
load-balances across a pool of single-GPU \texttt{hf} replicas or
tensor-parallel \texttt{vLLM} servers interchangeably; this is what makes the
$\{$HF, vLLM$\}$ throughput matrix (\Cref{tab:throughput}) a one-line backend
swap. Our inference measurements show that the vLLM serving path provides substantially higher throughput than the Hugging Face path.\looseness=-1

\subsection{Extensibility}
\label{subsec:extensibility}

\paragraph{New loss formulation.}
The training objective is an isolated function of the grouped scores; contrastive
CE and listwise-KL distillation are two branches over the same
\texttt{(n\_groups, G)} tensor (\Cref{subsec:impl}). A new objective\,---\,MarginMSE~\citep{hofstatter2021improvingefficientneuralranking},
RankNet~\citep{burges2005learning}, or listwise loss formulations~\citep{cao2007learningtorank}\,---\,is one more branch; the scoring, parallelism, and data
machinery are untouched:
\begin{lstlisting}[style=tevpy]
if loss_kind == "contrastive":
    loss = F.cross_entropy(grouped_scores, labels)  # positive: 0
elif loss_kind == "distill":
    loss = F.kl_div(
        student_lp, teacher_p, reduction="batchmean") * (s_temp ** 2)
# elif loss_kind == "my_loss":
#     loss = my_objective(grouped_scores, ...)  # new loss
\end{lstlisting}

\paragraph{New backbone / architecture.}
Models are loaded through the HF-config-driven weight bridge, so any
bridge-supported architecture trains with no reranker-side change. The one place
a new \emph{family} may need attention is LoRA targeting, which is a named
registry mapping roles to module patterns\,---\,the same mechanism we added to make
LoRA work on MoE expert layers (\Cref{sec:moe}). Each new model family adds a corresponding registry entry:
\begin{lstlisting}[style=tevpy]
LORA_TARGET_GROUPS = {
    "attn":        ("linear_qkv", "linear_proj"),
    "moe_experts": (
        "*.experts.*.linear_fc1", "*.experts.*.linear_fc2"),
    # "my_family_block": ("...module.name.patterns...",),  # add
}
\end{lstlisting}

\paragraph{New evaluation benchmark.}
Scoring is split from benchmark definition: the metric core scores a ranklist
against qrels with no benchmark knowledge, while a \texttt{Benchmark} registry
entry carries the dataset list, qrels loader, split convention, and table
layout. Adding a suite beyond BEIR-15 (MTEB~\citep{muennighoff-etal-2023-mteb}, BRIGHT~\citep{su2025bright}, a private set) is one
entry; the metric core, orchestrator, renderer, and backends are reused
unchanged:
\begin{lstlisting}[style=tevpy]
BENCHMARKS = {
    "beir15": Benchmark(
        datasets=..., load_qrels=...,
        split_for=..., summary_layout=...),
    # "my_bench": Benchmark(...),  # run --benchmark my_bench
}
\end{lstlisting}

\paragraph{New inference / serving backend.}
The scoring server is defined by one interface: a backend loads a model and
implements \texttt{score()} returning per-candidate scores; the HTTP layer,
client, load-balancing, and evaluator are backend-agnostic. Supporting a new
engine (e.g., SGLang~\citep{zheng2024sglang}, TensorRT-LLM)\footnote{\url{https://github.com/NVIDIA/TensorRT-LLM}} is a new subclass\,---\,exactly how the
\texttt{vllm} and \texttt{hf} backends already coexist behind the same contract:
\begin{lstlisting}[style=tevpy,columns=fixed,basewidth=0.5em]
class _BaseBackend:
    def score(self, req: ScoreRequest) -> list[ScoreItem]: ...

class VLLMBackend(_BaseBackend): ...   # implemented
class HFBackend(_BaseBackend): ...     # implemented
class MyEngineBackend(_BaseBackend):   # new backend
    def score(self, req): ...
\end{lstlisting}

\subsection{Backend Characterization}
\label{subsec:profiling}
Before introducing the Megatron backend, we characterize what the existing
PyTorch FSDP1 backend can and cannot do, and then compare with our new Megatron backend. We fine-tune the same dense 8B reranker
(Qwen3-8B~\citep{yang2025qwen3technicalreport}, full parameters, contrastive loss, RLHN-680K~\citep{thakur-etal-2025-hard}, one epoch, 64 queries
per step, single 8$\times$H200 node) under three data-parallel schemes and
measure wall-clock and peak GPU memory per rank (\Cref{tab:fsdp-profile}). Peak
memory is read from the training run's logged GPU statistics.

\paragraph{Profiling the existing FSDP backend} We make two observations. First, full-parameter 8B training requires
sharding: plain DDP, which replicates the optimizer state on every rank,
exhausts all 140 GiB and fails at the first optimizer step\,---\,an 8B model under
mixed-precision AdamW needs roughly 128 GiB of model states per rank (parameters,
gradients, fp32 master copy, and two moment buffers) before activations.
Second, among the training configurations that can fit in one 8$\times$H200 node, the heavier sharding option does not bring throughput improvement: \texttt{full\_shard} (ZeRO-3), which additionally shards parameters and
all-gathers them per layer, matches \texttt{shard\_grad\_op} (ZeRO-2) in both
wall-clock (11.4 vs $\sim$11.6\,h) and peak memory (51.2 vs 48.8 GiB),
because at 8B on one node the optimizer state\,---\,sharded in both\,---\,dominates the
footprint and the extra parameter gather is hidden by prefetch. The existing
backend is therefore \emph{memory-safe but throughput-bound}: it leaves no
obvious headroom to recover by tuning the sharding stage, which motivates a
backend with a faster data-parallel scheme.

\paragraph{Megatron matches FSDP quality and efficiency}
A faster backend is only useful if it does not sacrifice task performance. We verify
the training quality by training the dense 8B reranker with both backends under matched data,
recipe, and global batch, and evaluating on BEIR-15 (mean NDCG@10 over 15 tasks;
\Cref{subsec:setup}) across all three first-stage retrievers. \Cref{tab:backend-quality}
shows the two backends are indistinguishable on the full-parameter path: every
first stage agrees within $0.002$ NDCG@10. The LoRA path likewise tracks closely,
with the Hugging Face variant marginally ahead\,---\,by $0.008$ to $0.012$ NDCG@10, a gap comparable to the small differences in the loss comparison (\Cref{sec:ablation}) that we cannot reliably distinguish from the variation between independent training runs.

\begin{table}[t!]
\centering
\footnotesize
\setlength{\tabcolsep}{4pt}
\caption{Profiling the PyTorch FSDP1 backend on dense 8B full-parameter
fine-tuning, single 8$\times$H200 node, matched 64-query batch. Plain DDP
replicates all state; ZeRO-2 shards gradients and optimizer state; ZeRO-3 also
shards parameters. Plain DDP is
infeasible; the two sharded schemes fit but differ negligibly.\looseness=-1}
\begin{tabular}{llrrc}
\toprule
DP scheme & Sharding & Wall-clock & Peak mem/GPU & \makecell{Fits\\1 node} \\
\midrule
Plain DDP            & none     & \,---\,       & \makecell[r]{OOM\\($>$140 GiB)} & \xmark \\
FSDP \texttt{shard\_grad\_op} & ZeRO-2 & $\sim$11.6\,h & 48.8 GiB & \cmark \\
FSDP \texttt{full\_shard}     & ZeRO-3 & 11.4\,h & 51.2 GiB & \cmark \\
\bottomrule
\end{tabular}

\label{tab:fsdp-profile}
\end{table}

\begin{table}[t]
\centering
\footnotesize
\setlength{\tabcolsep}{4pt}
\caption{Reranker quality (BEIR-15 NDCG@10) is backend-invariant on the
full-parameter path (agreement within $0.002$ on every first stage). Dense 8B,
contrastive loss.}
\begin{tabular}{llccc}
\toprule
Backend & Param-eff. & BM25 & E5 & SPLADE-v3 \\
\midrule
\textit{First stage (no reranker)} & \,---\, & \textit{0.415} & \textit{0.503} & \textit{0.504} \\
\midrule
PyTorch FSDP1 & full-FT & 0.595 & 0.623 & 0.617 \\
Megatron           & full-FT & 0.596 & 0.624 & 0.619 \\
\addlinespace
PyTorch DDP & LoRA    & 0.598 & 0.626 & 0.620 \\
Megatron           & LoRA    & 0.590 & 0.614 & 0.610 \\
\bottomrule
\end{tabular}

\label{tab:backend-quality}
\end{table}

With quality controlled, we now turn to efficiency, and report it at two levels:
a \emph{configuration} comparison (each backend in its recommended setup) and a
\emph{matched-topology} comparison (same DP degree and activation recompute) that
isolates framework efficiency from the chosen topology.

\paragraph{Recommended-configuration comparison.}
Under each backend's recommended single-node configuration, the Megatron backend 
($\text{TP}=2, \text{DP}=4$, without activation recomputation) completes one 
dense-8B epoch in 8h\,55m compared to 11h\,22m for PyTorch FSDP1 \texttt{full\_shard}, 
a 22\% reduction in wall-clock time. A similar gap appears under LoRA 
(8h\,47m vs.\ 10h\,50m, $\sim$19\%). Because these default setups differ in both 
parallel topology and recomputation policies, this comparison reflects practical 
out-of-the-box runtimes rather than an isolated framework speedup.

\paragraph{Matched-topology comparison.}
To isolate the framework, we re-run both backends at \emph{data parallelism-matched} DP=8 with
activation recompute enabled on both: PyTorch FSDP1 \texttt{shard\_grad\_op}
(ZeRO-2) vs.\ Megatron TP=1/DP=8 (ZeRO-1) (\Cref{tab:backend-matched}). Here the
gap nearly closes\,---\,Megatron is only $\sim$4\% faster with
$\sim$2$\times$ the per-rank memory, because ZeRO-1 replicates parameters and
gradients while FSDP \texttt{shard\_grad\_op} shards gradients. These results demonstrate that the execution efficiency of both backends is comparable at matched topology. The observed speedup of the recommended Megatron setup primarily stems from operating in a lower-communication topology (TP=2, DP=4) without activation recomputation, rather than low-level runtime overheads. These results identify expert parallelism for MoE training (\Cref{sec:moe}), rather than dense throughput alone, as the main advantage of the Megatron backend. PyTorch FSDP1 provides no corresponding expert-parallel training path.

\begin{table}[t]
\centering
\caption{Dense-8B full-parameter training efficiency, single 8$\times$H200 node,
matched 64-query batch. \emph{Recommended} compares each backend's out-of-the-box
config (different topology + recompute); \emph{Matched DP=8} holds DP and
activation recompute fixed to isolate the framework, where the two are within
$\sim$4\% and Megatron's ZeRO-1 trades memory for the small speed edge. Megatron's
real differentiator is expert parallelism (\Cref{sec:moe}), not dense throughput.}
\footnotesize
\setlength{\tabcolsep}{4pt}
\begin{tabular}{lllrr}
\toprule
Comparison & Backend (topology) & Shard & Wall-clock & Mem/GPU \\
\midrule
\multirow{2}{*}{Recommended} & FSDP1 (DP=8) & ZeRO-3 & 11h\,22m & 51\,GiB \\
 & Megatron (TP=2/DP=4)        & ZeRO-1 & \textbf{8h\,55m}  & 106\,GiB \\
\addlinespace
\multirow{2}{*}{\makecell[l]{Matched\\DP=8}} & FSDP1 (DP=8) & ZeRO-2 & 11h\,00m & \textbf{49\,GiB} \\
 & Megatron (TP=1/DP=8)        & ZeRO-1 & \textbf{10h\,36m} & 106\,GiB \\
\bottomrule
\end{tabular}

\label{tab:backend-matched}
\end{table}

\begin{table*}[t]
\centering
\scriptsize
\setlength{\tabcolsep}{2.5pt}
\renewcommand{\arraystretch}{1.03}
\caption{NDCG@10 reranking results on different first-stage retrievers (top-200 candidates). Task abbreviations: ARG arguana, CFE climate-fever, DBP dbpedia-entity, FEV fever, FIQ fiqa, HOP hotpotqa, NFC nfcorpus, NQ nq, QUO quora, SCD scidocs, SCF scifact, TRC trec-covid, TOU webis-touche2020, MSM MS MARCO dev, CQA CQADupstack (avg of 12). Row labels: Dense 8B = Qwen3-8B, MoE 30B-A3B = Qwen3-30B-A3B; C = contrastive, D = listwise-KL distillation; F = full-parameter, L = LoRA.}
\resizebox{\textwidth}{!}{
\begin{tabular}{l*{16}{r}}
\toprule
Method & ARG & CFE & DBP & FEV & FIQ & HOP & NFC & NQ
& QUO & SCD & SCF & TRC & TOU & MSM & CQA & BEIR-15 \\
\midrule
\rowcolor{groupgray}\multicolumn{17}{l}{\textit{BM25}} \\

First & 0.397 & 0.165 & 0.318 & 0.651 & 0.236 & 0.633 & 0.338 & 0.306 & 0.789 & 0.149 & 0.679 & 0.595 & 0.442 & 0.228 & 0.302 & 0.415 \\
Dense 8B C/F & 0.797 & 0.361 & 0.487 & 0.911 & 0.512 & 0.833 & 0.410 & 0.658 & 0.810 & 0.261 & 0.793 & 0.887 & 0.353 & 0.427 & 0.447 & 0.596 \\
Dense 8B C/L & 0.734 & 0.318 & 0.504 & 0.903 & 0.499 & 0.828 & 0.407 & 0.656 & 0.837 & 0.247 & 0.798 & 0.885 & 0.358 & 0.424 & 0.445 & 0.590 \\
Dense 8B D/F & 0.733 & 0.411 & 0.481 & 0.909 & 0.485 & 0.810 & 0.404 & 0.623 & 0.780 & 0.242 & 0.795 & 0.865 & 0.424 & 0.433 & 0.391 & 0.586 \\
Dense 8B D/L & 0.686 & 0.392 & 0.493 & 0.907 & 0.472 & 0.809 & 0.396 & 0.624 & 0.826 & 0.236 & 0.792 & 0.854 & 0.410 & 0.431 & 0.406 & 0.582 \\
MoE 30B-A3B C/F & 0.809 & 0.374 & 0.494 & 0.908 & 0.520 & 0.834 & 0.410 & 0.662 & 0.829 & 0.271 & 0.809 & 0.882 & 0.323 & 0.433 & 0.450 & 0.600 \\
MoE 30B-A3B C/L & 0.747 & 0.321 & 0.497 & 0.896 & 0.494 & 0.826 & 0.411 & 0.659 & 0.854 & 0.261 & 0.794 & 0.895 & 0.288 & 0.426 & 0.455 & 0.588 \\
MoE 30B-A3B D/F & 0.739 & 0.408 & 0.483 & 0.910 & 0.490 & 0.810 & 0.406 & 0.624 & 0.793 & 0.247 & 0.795 & 0.858 & 0.400 & 0.433 & 0.402 & 0.587 \\
MoE 30B-A3B D/L & 0.661 & 0.389 & 0.490 & 0.905 & 0.468 & 0.809 & 0.404 & 0.623 & 0.841 & 0.240 & 0.790 & 0.858 & 0.403 & 0.429 & 0.382 & 0.579 \\
\midrule
\rowcolor{groupgray}\multicolumn{17}{l}{\textit{E5-base-v2}} \\

First & 0.446 & 0.266 & 0.422 & 0.850 & 0.399 & 0.691 & 0.354 & 0.582 & 0.861 & 0.187 & 0.720 & 0.696 & 0.264 & 0.427 & 0.374 & 0.503 \\
Dense 8B C/F & 0.801 & 0.406 & 0.526 & 0.944 & 0.581 & 0.867 & 0.427 & 0.744 & 0.790 & 0.278 & 0.809 & 0.890 & 0.331 & 0.474 & 0.487 & 0.624 \\
Dense 8B C/L & 0.739 & 0.344 & 0.542 & 0.935 & 0.555 & 0.860 & 0.422 & 0.743 & 0.825 & 0.264 & 0.816 & 0.888 & 0.330 & 0.468 & 0.482 & 0.614 \\
Dense 8B D/F & 0.743 & 0.479 & 0.513 & 0.941 & 0.551 & 0.839 & 0.419 & 0.700 & 0.770 & 0.256 & 0.807 & 0.883 & 0.408 & 0.478 & 0.419 & 0.614 \\
Dense 8B D/L & 0.694 & 0.453 & 0.529 & 0.939 & 0.531 & 0.836 & 0.402 & 0.699 & 0.821 & 0.250 & 0.804 & 0.876 & 0.399 & 0.479 & 0.434 & 0.610 \\
MoE 30B-A3B C/F & 0.815 & 0.430 & 0.530 & 0.941 & 0.599 & 0.868 & 0.431 & 0.751 & 0.814 & 0.289 & 0.827 & 0.880 & 0.305 & 0.476 & 0.490 & 0.630 \\
MoE 30B-A3B C/L & 0.756 & 0.348 & 0.541 & 0.926 & 0.562 & 0.859 & 0.434 & 0.746 & 0.846 & 0.277 & 0.808 & 0.887 & 0.271 & 0.470 & 0.496 & 0.615 \\
MoE 30B-A3B D/F & 0.750 & 0.475 & 0.514 & 0.942 & 0.553 & 0.840 & 0.421 & 0.698 & 0.785 & 0.259 & 0.806 & 0.876 & 0.390 & 0.478 & 0.429 & 0.615 \\
MoE 30B-A3B D/L & 0.669 & 0.452 & 0.520 & 0.935 & 0.528 & 0.837 & 0.416 & 0.697 & 0.836 & 0.250 & 0.804 & 0.874 & 0.397 & 0.475 & 0.407 & 0.606 \\

\midrule
\rowcolor{groupgray}\multicolumn{17}{l}{\textit{SPLADE-v3}} \\

First & 0.488 & 0.256 & 0.445 & 0.810 & 0.380 & 0.689 & 0.363 & 0.586 & 0.814 & 0.156 & 0.716 & 0.732 & 0.312 & 0.459 & 0.344 & 0.504 \\
Dense 8B C/F & 0.815 & 0.391 & 0.524 & 0.939 & 0.561 & 0.831 & 0.416 & 0.742 & 0.798 & 0.270 & 0.807 & 0.899 & 0.349 & 0.471 & 0.471 & 0.619 \\
Dense 8B C/L & 0.748 & 0.338 & 0.546 & 0.930 & 0.543 & 0.826 & 0.408 & 0.740 & 0.829 & 0.256 & 0.812 & 0.892 & 0.348 & 0.466 & 0.467 & 0.610 \\
Dense 8B D/F & 0.749 & 0.460 & 0.518 & 0.936 & 0.528 & 0.807 & 0.410 & 0.697 & 0.771 & 0.249 & 0.805 & 0.886 & 0.427 & 0.475 & 0.407 & 0.608 \\
Dense 8B D/L & 0.697 & 0.434 & 0.532 & 0.935 & 0.508 & 0.805 & 0.393 & 0.695 & 0.821 & 0.244 & 0.802 & 0.869 & 0.409 & 0.476 & 0.423 & 0.603 \\
MoE 30B-A3B C/F & 0.828 & 0.406 & 0.530 & 0.936 & 0.581 & 0.833 & 0.418 & 0.749 & 0.819 & 0.279 & 0.823 & 0.885 & 0.319 & 0.473 & 0.473 & 0.623 \\
MoE 30B-A3B C/L & 0.763 & 0.342 & 0.540 & 0.923 & 0.537 & 0.824 & 0.416 & 0.744 & 0.849 & 0.266 & 0.804 & 0.900 & 0.278 & 0.467 & 0.478 & 0.608 \\
MoE 30B-A3B D/F & 0.756 & 0.456 & 0.519 & 0.938 & 0.532 & 0.807 & 0.414 & 0.696 & 0.783 & 0.253 & 0.804 & 0.882 & 0.392 & 0.476 & 0.418 & 0.608 \\
MoE 30B-A3B D/L & 0.672 & 0.431 & 0.524 & 0.931 & 0.505 & 0.806 & 0.409 & 0.695 & 0.834 & 0.245 & 0.800 & 0.866 & 0.400 & 0.474 & 0.396 & 0.599 \\
\bottomrule
\end{tabular}
}
\label{tab:moe-vs-dense}
\end{table*}

The backend supports LoRA and full-parameter fine-tuning identically. We turn to expert parallelism next.

\section{Expert-Parallel MoE Training}
\label{sec:moe}

The key advantage of the Megatron backend is that it enables mixture-of-experts training.
A modern MoE reranker like Qwen3-30B-A3B~\citep{yang2025qwen3technicalreport} carries 30B total parameters across 128 experts but activates only $\sim$3B per token (top-8 routing). The memory bottleneck is from the \emph{experts}, not the attention/feed-forward: holding all 128 experts' weights and optimizer state on every rank is inefficient or infeasible. PyTorch FSDP1 can only replicate-then-shard them as ordinary parameters, with no mechanism for routing computation to the experts on a given rank.

\paragraph{Expert parallelism and topology}
Megatron introduces expert parallelism (EP) as an additional parallel dimension alongside the tensor- and data-parallel axes described in \Cref{subsec:impl}. Whereas TP shards the dense transformer layers and DP replicates the model across data-parallel ranks, EP partitions only the MoE experts. In practice, the EP dimension folds into the existing 3D Megatron topology by subdividing each data-parallel group into expert-parallel subgroups: dense layers continue to execute under the same TP$\times$DP configuration, while MoE layers dispatch routed tokens only among the ranks within an EP group. As a result, the overall world size is still TP$\times$DP (with pipeline parallelism omitted here), with EP dividing DP in our configuration: the experts of each MoE layer are partitioned $\mathrm{EP}$ ways and replicated $\mathrm{DP}/\mathrm{EP}$ times, and the distributed optimizer shards their optimizer state only across those $\mathrm{DP}/\mathrm{EP}$ replicas, while the dense layers and their optimizer-state shards span all $\mathrm{DP}$ ranks, and only the MoE layers incur the additional expert communication. Per-rank expert memory therefore scales with $\text{experts}/\mathrm{EP}$ rather than the full expert count. Training Qwen3-30B-A3B across two 8$\times$H200 nodes, i.e., sixteen GPUs with $\mathrm{TP}{=}1$, $\mathrm{DP}{=}16$, and $\mathrm{EP}{=}16$, so that each GPU holds one sixteenth of the experts, with an expert replication factor $\mathrm{DP}/\mathrm{EP}$ of one and hence the full optimizer state of its own experts, peaks at 78.9 GiB per GPU because the 128 experts are partitioned sixteen ways, a capability unavailable in PyTorch FSDP1.

\paragraph{LoRA target groups for MoE}
Applying LoRA to an MoE model exposes a subtlety. The conventional dense target
specification adapts the attention and MLP projection modules; but on an MoE
model the MLP leaf modules (\texttt{linear\_fc1}/\texttt{linear\_fc2}) also name
every one of the 128 expert feed-forward networks, so a naive dense pattern
either silently adapts all experts (an enormous, unintended parameter count) or,
if restricted to attention only, leaves the experts\,---\,where the reranking signal
concentrates\,---\,untouched. We observe the latter failure directly: an
attention-only LoRA on the MoE backbone underperforms its full-parameter
counterpart by a wide margin, worse even than dense LoRA at far smaller scale. We
therefore expose a named target-group registry (\Cref{subsec:impl}) that maps roles
(\texttt{attn}, \texttt{moe\_experts}, \texttt{moe\_shared}, \texttt{moe\_router})
to the correct module patterns, and use the expert-aware specification
(\texttt{attn}\,+\,\texttt{moe\_experts}\,+\,\texttt{moe\_shared}, leaving the
router frozen for stability) for all MoE LoRA runs in our study.

\begin{table}[t]
\centering
\footnotesize
\setlength{\tabcolsep}{4pt}
\caption{Reranking throughput (pairs/s, summed over eight independent single-GPU workers on H200s; BM25 candidate
set; contrastive full-FT checkpoints). The MoE serves faster than dense 8B, and
the advantage grows under vLLM.}
\begin{tabular}{lccc}
\toprule
Backend & 8B dense & 30B-A3B MoE & MoE speedup \\
\midrule
Hugging Face & 579 & 668 & $1.15\times$ \\
vLLM        & 1{,}834 & 2{,}622 & $1.43\times$ \\
\midrule
vLLM speedup & $3.17\times$ & $3.93\times$ & \\
\bottomrule
\end{tabular}

\label{tab:throughput}
\end{table}

\begin{table*}[t]
\centering
\caption{Sanity check of our released checkpoints against prior
rerankers (per-dataset NDCG@10 on 15 BEIR tasks, all reranking SPLADE-v3 top-200). Baselines differ in training data and training-stage retrievers. The results place our released checkpoints in the expected range. AVG reports the
BEIR-13 mean, excluding CQADupstack and MS MARCO, as commonly reported by prior works. One contrastive and one distilled full-parameter checkpoint per backbone. Best result per column in bold, second best underlined; the first-stage row is excluded.}
\vspace{0pt}
\resizebox{\textwidth}{!}{%
\begin{tabular}{l rrrrrrrrrrrrrrr r}
\toprule
Model & ARG & CFE & DBP & FEV & FIQ & HOP & NFC & NQ\phantom{0} & QUO & SCD & SCF & TRC & TOU & MSM & CQA & AVG \\
\midrule
\rowcolor{groupgray}\multicolumn{17}{l}{\textit{First stage}} \\
SPLADE-v3 (no rerank) & 0.488 & 0.256 & 0.445 & 0.810 & 0.380 & 0.689 & 0.363 & 0.586 & 0.814 & 0.156 & 0.716 & 0.732 & 0.312 & 0.459 & 0.344 & 0.519 \\
\midrule
\rowcolor{groupgray}\multicolumn{17}{l}{\textit{Prior rerankers}} \\
RankT5~\citep{zhuang2023rankt5} & 0.330 & 0.215 & 0.442 & 0.832 & 0.445 & 0.710 & 0.381 & 0.614 & 0.831 & 0.181 & 0.750 & 0.807 & \textbf{0.440} & - & - & 0.537 \\
RankLLaMA~\citep{ma2024fine} & 0.560 & 0.280 & 0.483 & 0.839 & 0.465 & 0.753 & 0.303 & 0.663 & 0.850 & 0.178 & 0.732 & 0.852 & 0.401 & - & - & 0.566 \\
RankQwen~\citep{xu2025distillationversuscontrastivelearning} & 0.791 & 0.405 & \textbf{0.543} & \textbf{0.940} & 0.559 & \textbf{0.848} & \textbf{0.424} & \underline{0.745} & 0.776 & \underline{0.271} & \underline{0.819} & 0.883 & 0.329 & - & - & 0.641 \\
ModernBERT Reranker~\citep{déjean2025rerankingcompresseddocumentrepresentation} & - & 0.253 & 0.501 & 0.884 & 0.476 & 0.754 & 0.381 & 0.660 & 0.860 & 0.192 & 0.754 & \underline{0.890} & 0.352 & - & - & -\\
Mistral-7B Reranker~\citep{déjean2025rerankingcompresseddocumentrepresentation} & - & 0.275 & 0.505 & 0.869 & 0.482 & 0.741 & 0.385 & 0.672 & \underline{0.864} & 0.192 & 0.773 & 0.856 & 0.316 & - & - & -\\
RRK~\citep{déjean2025rerankingcompresseddocumentrepresentation} & - & 0.259 & 0.494 & 0.842 & 0.471 & 0.734 & 0.387 & 0.668 & \textbf{0.870} & 0.195 & 0.764 & 0.862 & 0.289 & - & - & -\\

\midrule
\rowcolor{groupgray}\multicolumn{17}{l}{\textit{Ours (Tevatron~3.0)}} \\
Dense 8B, contrastive   & \underline{0.815} & 0.391 & 0.524 & \underline{0.939} & \underline{0.561} & 0.831 & 0.416 & 0.742 & 0.798 & 0.270 & 0.807 & \textbf{0.899} & 0.349 & 0.471 & \underline{0.471} & \underline{0.642} \\
Dense 8B, distill       & 0.749 & \textbf{0.460} & 0.518 & 0.936 & 0.528 & 0.807 & 0.410 & 0.697 & 0.771 & 0.249 & 0.805 & 0.886 & \underline{0.427} & \underline{0.475} & 0.407 & 0.634 \\
MoE 30B-A3B, contrastive & \textbf{0.828} & 0.406 & \underline{0.530} & 0.936 & \textbf{0.581} & \underline{0.833} & \underline{0.418} & \textbf{0.749} & 0.819 & \textbf{0.279} & \textbf{0.823} & 0.885 & 0.319 & 0.473 & \textbf{0.473} & \textbf{0.646} \\
MoE 30B-A3B, distill     & 0.756 & \underline{0.456} & 0.519 & 0.938 & 0.532 & 0.807 & 0.414 & 0.696 & 0.783 & 0.253 & 0.804 & 0.882 & 0.392 & \textbf{0.476} & 0.418 & 0.633 \\
\bottomrule
\end{tabular}}

\label{tab:main-comparison}
\vspace{-3pt}
\end{table*}

\section{Demonstration: A Controlled Study of Reranker Recipes}
\label{sec:study}

Having established the backend (\Cref{sec:backend}) and its MoE capability
(\Cref{sec:moe}), we use Tevatron 3.0 to run a study that would be awkward or infeasible on
the prior stack: a controlled comparison of model architecture, parameter efficiency,
and loss formulation, all trained through one backend and evaluated through one
pipeline. We present it as a \emph{demonstration of the system}\,---\,evidence that
the framework supports controlled, apples-to-apples comparison at scale, and release the checkpoints for reproducibility.

\subsection{Setup}
\label{subsec:setup}
All rerankers are trained on RLHN-680K~\citep{thakur-etal-2025-hard} for one epoch with a listwise group size
of 8 (one positive, seven hard negatives), maximum sequence length 512, bf16, on
a single 8$\times$H200 node (two nodes for the MoE runs), holding the global batch
at 64 queries per step so that throughput and wall-clock are directly comparable.
Full-parameter and LoRA runs share this recipe; LoRA uses a $10\times$ larger
learning rate. Distillation runs minimize a listwise KL against per-candidate
teacher scores precomputed offline from a Qwen3-Reranker-8B teacher~\citep{zhang2025qwen3embedding}. 
We report one training run per configuration. Without repeated runs, we cannot determine whether small performance differences reflect systematic effects or random variation across independent runs.

We evaluate on BEIR-15~\citep{thakur2021beir}\,---\,the mean NDCG@10 over 15 tasks (BEIR-13, MS
MARCO dev, and CQADupstack averaged over its 12 sub-forums)\,---\,reranking the top
200 candidates of each first stage. We use three first-stage retrievers chosen to
span the lexical / dense / learned-sparse spectrum \emph{and} a range of first-stage
strength: BM25~\citep{robertson1995okapi}, E5-base-v2~\citep{wang2022textembeddingbyweaklysupervisedpretraining}, and SPLADE-v3~\citep{lassance2024spladev3}, with BEIR-15 NDCG@10 of 0.415, 0.503,
and 0.504 respectively. We deliberately avoid first stages strong enough to leave
no reranking headroom; reporting all three lets us read how the reranking gain
varies with first-stage quality rather than reading a single operating point.

\subsection{MoE Quality and Serving Throughput}
\label{subsec:moe-vs-dense}

The central question for deployment is whether the MoE reranker, activating less
than half the parameters per token, gives up quality relative to the dense 8B.
As shown in~\cref{tab:moe-vs-dense}, the 30B-A3B MoE model matches the retrieval effectiveness of the dense 8B baseline across all first-stage retrievers, despite activating fewer than half the parameters per token.
The aggregate occasionally favors the MoE, but the margins are small, with no variance estimate to judge them against, and the macro-average hides per-task structure, so we do not claim the MoE is strictly better performance-wise.

\paragraph{Inference throughput.}
We next test whether fewer activated parameters translate into higher serving throughput. \Cref{tab:throughput} measures reranking throughput (query--passage pairs
per second) over the full BM25 candidate set, scoring with eight independent
single-GPU workers under two serving backends, Hugging Face and vLLM. Two effects
are visible and they compound. Across backends, vLLM is $3.2$--$3.9\times$ faster
than Hugging Face, consistent with its use of paged-KV and continuous batching~\citep{kwon2023efficientmemorymanagement}. Across
architectures, the MoE is faster than the dense 8B\,---\,$1.15\times$ on Hugging Face,
widening to $1.43\times$ on vLLM. The MoE's smaller active parameter count and vLLM's batching are plausible contributors to these gains. The matrix measures end-to-end throughput under each configuration; isolating the contributions of batching, kernels, and expert routing would require additional profiling. The fastest configuration (vLLM\,$\times$\,MoE,
2{,}622 pairs/s) delivers $4.5\times$ the throughput of the slowest
(Hugging Face\,$\times$\,8B, 579 pairs/s).

The 30B-A3B MoE thus matches dense-8B quality with higher serving throughput in this setting. Training it uses two nodes with expert parallelism, versus one for the dense 8B; total training cost and a serving break-even point remain unmeasured.

\subsection{Parameter Efficiency and Distilled Training}
\label{sec:ablation}

\paragraph{LoRA recovers most of full fine-tuning.}
In \Cref{tab:moe-vs-dense}, rank-16 LoRA on the dense 8B trails full-parameter training by
$\sim$0.006--0.010 NDCG@10 across first stages while training a small fraction of
the parameters; on the MoE, the expert-aware target specification
(\Cref{sec:moe}) is what makes this recovery possible at all. Combined with the
near-identical wall-clock of LoRA and full-parameter runs (the frozen-base
forward/backward dominates), LoRA primarily provides memory savings rather than latency reductions, offering a practical trade-off between peak memory and ranking accuracy.

\paragraph{Distillation and contrastive learning specialize on different datasets.}
Across the evaluated configurations, contrastive training achieves higher mean BEIR-15 NDCG@10 than distillation, although distillation improves performance on some individual datasets. Per-task analysis reveals that the aggregate BEIR-15 score obscures substantial dataset-specific performance divergence. With the teacher temperature fixed at 2, distillation improves NDCG@10 on CFE and TOU but reduces it on ARG and NQ across the evaluated configurations. We believe two structural factors bound the effect further:
our 8B teacher is the \emph{same size} as the dense student and, for the 30B-A3B MoE student, smaller in total but larger in active parameters, so neither student has a clearly larger teacher to distill from, and the training data's hard negatives are
already denoised, removing distillation's usual false-negative-relabeling value.
We therefore report the loss comparison as \emph{illustrative under this fixed
recipe}; pinning down the exact contrastive-vs-distillation tradeoff\,---\,its
dependence on temperature, group size, and teacher capacity\,---\,is beyond this
work's scope. 

\subsection{Sanity Check Against Prior Rerankers}
\label{sec:baselines}

We include a comparison against prior reranker baselines (\Cref{tab:main-comparison})
as a \emph{sanity check on our training pipeline}, not as a competitive claim.
Tevatron~3.0 is a training-and-serving system; the checkpoints in this paper are
artifacts produced \emph{by} that system under one fixed recipe, released so that
others can reproduce and build on them. Because these checkpoints are reproducible baseline artifacts rather than heavily tuned competitive submissions, small differences should not be over-interpreted. The
baselines were also generally trained on different data with different first stages,
so the rows are not strictly controlled.
All rows rerank the same SPLADE-v3 top-200 candidates; the first row is that
first stage alone. Per-dataset NDCG@10 uses the task abbreviations of \Cref{tab:moe-vs-dense}, and AVG is the BEIR-13 mean as commonly reported by prior works~\citep{zhuang2023rankt5,ma2024fine,xu2025distillationversuscontrastivelearning,Zhuang2026LayerwiseTC}.

The per-dataset results against the strongest baseline, RankQwen, are mixed in both directions and the margins are small. Against our contrastive MoE checkpoint, the row with the highest mean, RankQwen leads on five datasets (DBP, FEV, HOP, NFC, and TOU) by $0.004$ to $0.015$ NDCG@10 and trails on the other eight by $0.001$ to $0.043$, and the two BEIR-13 means differ by $0.005$. With single runs and no variance estimates (\Cref{subsec:setup}), we do not read this as an advantage in either direction.
A system-level comparison would require retraining the baselines under our recipe, which the backend makes possible but which is outside the scope of this paper.

\section{Conclusion and Future Work}
\label{sec:conclusion}

We present Tevatron~3.0, which integrates a Megatron-Core training backend into 
the Tevatron reranker toolkit while preserving its existing data processing, evaluation 
pipeline, and Hugging Face checkpoint interoperability. Profiling the existing PyTorch FSDP1 
path demonstrates that it is memory-safe but throughput-bound and lacks expert parallelism. 
The Megatron backend matches FSDP1 reranker quality and, at matched data-parallel topology, 
its execution efficiency (while reducing epoch time by $\sim$22\% in its recommended 
configuration). Crucially, expert parallelism makes 30B-scale MoE reranker training feasible 
on a two-node academic budget. Our controlled study demonstrates that a 30B-A3B MoE reranker 
matches dense-8B retrieval effectiveness with fewer than half the activated parameters and 
substantially higher serving throughput, LoRA recovers most full-parameter quality, and 
distillation helps on some datasets but lowers the BEIR-15 mean under our fixed recipe. We release all framework code 
and trained checkpoints as reusable artifacts for the community to continue building on Tevatron~3.0.

In the broader IR ecosystem, Tevatron~3.0 addresses a persistent infrastructure gap. 
While established neural retrieval toolkits (such as OpenMatch~\citep{liu2021openmatchopensourcelibrary} and FlagEmbedding~\citep{bge_embedding}) support 
standard bi-encoder and cross-encoder pipelines, they rely on trainers that lack native 
expert parallelism. Concurrently, recent work has begun exploring sparse MoE architectures 
for dense retrieval embeddings~\cite{nussbaum2025trainingsparsemixtureexperts,shrestha2026mixtureofexpertslanguagemodelsstrong}. Tevatron~3.0 
complements this line of research by providing a reliable and robust open-source, expert-parallel training 
and serving pipeline specifically engineered for generative MoE cross-encoders with backward compatibility.

Several directions remain open. The distillation comparison calls for
experiments varying temperature, group size, and\,---\,most promisingly\,---\,teacher capacity,
since our 8B teacher leaves no clear capacity headroom to distill into. Broader
architecture and scale coverage, and extending the backend's pipeline and context
parallelism to longer-context reranking, are natural next steps. Three limitations of the present study also mark out next steps. First, the compute we call an academic budget, one 8-GPU node for the dense runs and two for the MoE run, is still beyond many groups; tensor and pipeline parallelism, activation recompute, and LoRA are the backend's levers for smaller or cheaper allocations, but we have not profiled them on GPUs with less memory than an H200, and doing so would give practitioners concrete recipes for 80\,GB and smaller cards. Second, the systems comparison covers only the Hugging Face Trainer path of Tevatron~2.0, namely FSDP1 and, for LoRA, DDP; FSDP2, DeepSpeed ZeRO with its recently merged gradient-accumulation fix, and other Megatron-based training frameworks are not compared. Third, every result is a single run, and the comparisons in this paper should be repeated over several seeds before small differences are acted upon.

\newcounter{balancebib}
\let\oldbibitem\bibitem
\def\bibitem{\stepcounter{balancebib}\ifnum\value{balancebib}=37 \vfill\eject\fi\oldbibitem}

\bibliographystyle{ACM-Reference-Format}
\bibliography{technical-report/references}

\end{document}